\documentclass[preprint,nopreprintnumbers,titlepage,amsmath,amssymb,aps,pre]{revtex4-2}

\usepackage{graphicx}
\usepackage{dcolumn}
\usepackage{bm}
\usepackage{hyperref}

\usepackage{braket}
\usepackage{lipsum}
\usepackage{alltt}
\usepackage{multirow}
\usepackage{stackrel}
\usepackage{longtable}
\usepackage{lipsum}

\makeatletter
\allowdisplaybreaks[4]

\newcommand{\e}{{\rm e}}
\newcommand{\tr}{{\mathrm{tr}}}

\newcommand{\dd}{\,\mathrm{d}}

\begin{document}
\title{Exact variance of von Neumann entanglement entropy \\ over the Bures-Hall measure}
\author{Lu Wei}
\email{luwe@umich.edu}
\affiliation{Department of Electrical and Computer Engineering \\ University of Michigan - Dearborn, MI 48128, USA}
\date{\today}

\begin{abstract}
The Bures-Hall distance metric between quantum states is a unique measure that satisfies various useful properties for quantum information processing. In this work, we study the statistical behavior of quantum entanglement over the Bures-Hall ensemble as measured by von Neumann entropy. The average von Neumann entropy over such an ensemble has been recently obtained, whereas the main result of this work is an explicit expression of the corresponding variance that specifies the fluctuation around its average. The starting point of the calculations is the connection between correlation functions of the Bures-Hall ensemble and these of the Cauchy-Laguerre ensemble. The derived variance formula, together with the known mean formula, leads to a simple but accurate Gaussian approximation to the distribution of von Neumann entropy of finite-size systems. This Gaussian approximation is also conjectured to be the limiting distribution for large dimensional systems.
\end{abstract}

\maketitle

\section{Introduction and the main result}
Quantum information theory aims at understanding the theoretical underpinnings of quantum technologies such as quantum computing and quantum communications. Crucial to successful exploitation of the quantum revolutionary advances is the understanding of the non-classical phenomenon of quantum entanglement. Entanglement is the most fundamental characteristic trait of quantum mechanics, which is also the resource and medium that enable quantum technologies.

In this work, we aim to understand the statistical behavior of entanglement of quantum bipartite systems over Bures-Hall measure~\cite{Hall98,Zyczkowski01}. The Bures-Hall measure enjoys the property that, without any prior knowledge on a density matrix, the optimal way to estimate the density matrix is to generate a state at random with respect to this measure~\cite{Zyczkowski01,Sommers04,Osipov10}. In particular, we study the degree of entanglement as measured by the von Neumann entropy over such a measure. The mean value of von Neumann entropy over the Bures-Hall measure has been recently obtained~\cite{Sarkar19,Wei20a}. As an important step towards understanding its statistical distribution, we derive the corresponding variance in this paper. The variance describes the fluctuation of the entropy around its mean value, which also provides crucial information such as whether the average entropy is typical.

The density matrix formulism~\footnote{For a comprehensive treatment of the density matrix formulism, we refer readers to Refs.~\cite{BZ17,Majumdar} and references therein.}, introduced by von Neumann, that has led to the Bures-Hall ensemble is described as follows. Consider a composite (bipartite) system that consists of two subsystems $A$ and $B$ of Hilbert space dimensions $m$ and $n$, respectively. The Hilbert space $\mathcal{H}_{A+B}$ of the composite system is given by the tensor product of the subsystems, $\mathcal{H}_{A+B}=\mathcal{H}_{A}\otimes\mathcal{H}_{B}$. A random pure state of the composite system $\mathcal{H}_{A+B}$ is defined as a linear combination of the random coefficients $z_{i,j}$ and the complete basis $\left\{\Ket{i^{A}}\right\}$ and $\left\{\Ket{j^{B}}\right\}$ of $\mathcal{H}_{A}$ and $\mathcal{H}_{B}$~\cite{Majumdar},
\begin{equation}\label{eq:S0}
\Ket{\psi}=\sum_{i=1}^{m}\sum_{j=1}^{n}z_{i,j}\Ket{i^{A}}\otimes\Ket{j^{B}},
\end{equation}
where each $z_{i,j}$ follows the standard complex Gaussian distribution. We now consider a superposition of the state~(\ref{eq:S0}),
\begin{equation}\label{eq:SB}
\Ket{\varphi}=\Ket{\psi}+\left(\mathbf{U}\otimes\mathbf{I}_{m}\right)\Ket{\psi},
\end{equation}
where $\mathbf{U}$ is an $m\times m$ unitary random matrix with the measure proportional to $\det\left(\mathbf{I}_{m}+\mathbf{U}\right)^{2\alpha+1}$~\cite{Sarkar19}. The corresponding density matrix of the pure state~(\ref{eq:SB}) is
\begin{equation}\label{eq:rho}
\rho=\Ket{\varphi}\Bra{\varphi},
\end{equation}
which has the natural probability constraint
\begin{equation}\label{eq:del}
\tr(\rho)=1.
\end{equation}
We assume without loss of generality that $m\leq n$. The reduced density matrix $\rho_{A}$ of the smaller subsystem $A$ is computed by partial tracing (purification) of the full density matrix~(\ref{eq:rho}) over the other subsystem $B$ (environment) as
\begin{equation}\label{eq:rhoB}
\rho_{A}=\tr_{B}\rho.
\end{equation}
The resulting density of eigenvalues of $\rho_{A}$ ($\lambda_{i}\in[0,1]$, $i=1,\dots,m$) is the (generalized) Bures-Hall measure~\cite{Hall98,Zyczkowski01,Sarkar19}
\begin{equation}\label{eq:BH}
f\left(\bm{\lambda}\right)=\frac{1}{C}~\delta\left(1-\sum_{i=1}^{m}\lambda_{i}\right)\prod_{1\leq i<j\leq m}\frac{\left(\lambda_{i}-\lambda_{j}\right)^{2}}{\lambda_{i}+\lambda_{j}}\prod_{i=1}^{m}\lambda_{i}^{\alpha},
\end{equation}
where the parameter $\alpha$ takes half-integer values
\begin{equation}\label{eq:aBH}
\alpha=n-m-\frac{1}{2},
\end{equation}
and the constant $C$ is
\begin{equation}\label{eq:cBH}
C=\frac{2^{-m(m+2\alpha)}\pi^{m/2}}{\Gamma\left(m(m+2\alpha+1)/2\right)}\prod_{i=1}^{m}\frac{\Gamma(i+1)\Gamma(i+2\alpha+1)}{\Gamma(i+\alpha+1/2)}.
\end{equation}
In Eq.~(\ref{eq:BH}), the presence of the Dirac delta function $\delta(\cdot)$ reflects the constraint~(\ref{eq:del}). Note that another approach to obtain the measure~(\ref{eq:BH}) is by introducing a distance metric (Bures-Hall metric) over reduced density matrices. The Bures-Hall metric is the only monotone metric that is simultaneously Fisher adjusted and Fubini-Study adjusted~\cite{Zyczkowski01,Sommers04}. It is also a function of fidelity~\cite{BZ17}, which is a key performance indicator in quantum information processing.

The above bipartite model is useful in describing the entanglement between the two subsystems of different bipartite systems, in which one subsystem represents a physical object (such as spins) and the other subsystem is the environment (such as a heat bath). The degree of entanglement of subsystems can be measured by entanglement entropies, which are functions of eigenvalues (entanglement spectrum) of the reduced density matrix. We consider the standard measure of von Neumann entropy of the subsystem~\footnote{Note that since the composite system is in a random pure state, the von Neumann entropy of the full system is zero~\cite{Page93}.}
\begin{equation}\label{eq:vN}
S=-\tr\left(\rho_{A}\ln\rho_{A}\right)=-\sum_{i=1}^{m}\lambda_{i}\ln\lambda_{i},
\end{equation}
supported in $S\in\left[0,\ln{m}\right]$, which achieves the separable state ($S=0$) when $\lambda_{1}=1$, $\lambda_{2}=\dots=\lambda_{m}=0$ and the maximally-entangled state ($S=\ln{m}$) when $\lambda_{1}=\dots=\lambda_{m}=1/m$. Statistical information of entropies is encoded through the moments. In particular, the first moment (average value) implies the typical behavior of entanglement and the second moment (variance) specifies the fluctuation around the typical value. The average von Neumann entropy, valid for any subsystem dimensions $m\leq n$, has been recently obtained as~\cite{Sarkar19,Wei20a}
\begin{equation}\label{eq:vNm}
\mathbb{E}_{f}\!\left[S\right]=\psi_{0}\left(mn-\frac{m^2}{2}+1\right)-\psi_{0}\left(n+\frac{1}{2}\right),
\end{equation}
where the expectation $\mathbb{E}_{f}\!\left[\cdot\right]$ is taken over the Bures-Hall ensemble~(\ref{eq:BH}). Here,
$\psi_{0}(x)=\dd\ln\Gamma(x)/\dd x$ is the digamma function~\cite{Prudnikov86} and for a positive integer $l$,
\begin{subequations}
\begin{eqnarray}
\psi_{0}(l)&=&-\gamma+\sum_{k=1}^{l-1}\frac{1}{k},\\
\psi_{0}\left(l+\frac{1}{2}\right)&=&-\gamma-2\ln2+2\sum_{k=0}^{l-1}\frac{1}{2k+1},
\end{eqnarray}
\end{subequations}
where $\gamma\approx0.5772$ is the Euler's constant. The main result of this work is the following formula, proved in Sec.~\ref{sec:vNv}, of the exact variance of von Neumann entropy over the Bures-Hall ensemble
\begin{equation}\label{eq:vNv}
\mathbb{V}\!_{f}\!\left[S\right]=-\psi_{1}\left(mn-\frac{m^2}{2}+1\right)+\frac{2n(2n+m)-m^{2}+1}{2n(2mn-m^2+2)}\psi_{1}\left(n+\frac{1}{2}\right),
\end{equation}
where $\psi_{1}(x)=\dd^{2}\ln\Gamma(x)/\dd x^{2}$ is the trigamma function~\cite{Prudnikov86} and for a positive integer $l$,
\begin{equation}\label{eq:tg}
\psi_{1}(l)=\frac{\pi^{2}}{6}-\sum_{k=1}^{l-1}\frac{1}{k^{2}}.
\end{equation}
Under the Bures-Hall measure, other entropies such as the quantum purity have also been studied in the literature. In particular, the first a few exact moments of quantum purity~\cite{Sommers04,Osipov10,Sarkar19,Wei20a} as well as its asymptotic distribution~\cite{Nadal11} are known~\footnote{The results~\cite{Sommers04,Osipov10} are valid for the special case $m=n$.}. Besides the Bures-Hall measure, the exact moments of von Neumann entropy~\cite{Page93,Foong94,Ruiz95,VPO16,Wei17,Wei20} and quantum purity~\cite{Lubkin78,Giraud07} have been well-investigated over the arguably less complicated~\footnote{The Hilbert-Schmidt measure corresponds to the density without the interaction terms $\prod_{1\leq i<j\leq m}(\lambda_{i}+\lambda_{j})$ in~(\ref{eq:BH}).} Hilbert-Schmidt measure~\cite{BZ17}. Finally, we note that tools from asymptotic geometric analysis have also been utilized to characterize entanglement of large dimensional quantum systems~\cite{ASY14,AS17}.

With the expressions of the mean~(\ref{eq:vNm}) and variance~(\ref{eq:vNv}), simple approximations can be constructed to understand the distribution of von Neumann entropy. For convenience, we standardize the von Neumann entropy as
\begin{equation}\label{eq:X}
X=\frac{S-\mathbb{E}_{f}\!\left[S\right]}{\sqrt{\mathbb{V}\!_{f}\!\left[S\right]}}
\end{equation}
so that the random variable $X$, supported in $X\in(-\infty,\infty)$, has zero mean and unit variance. Thus, a natural approximation to the distribution of $X$ would be a standard Gaussian distribution
\begin{equation}\label{eq:Gau}
\varphi_{X}(x)=\frac{1}{\sqrt{2\pi}}\e^{-\frac{1}{2}x^{2}},
\end{equation}
i.e., the distribution of $S$ is approximated by a Gaussian distribution with mean $\mathbb{E}_{f}\!\left[S\right]$ and variance $\mathbb{V}\!_{f}\!\left[S\right]$.
\begin{figure}[!h]
\centering
\includegraphics[width=0.7\linewidth]{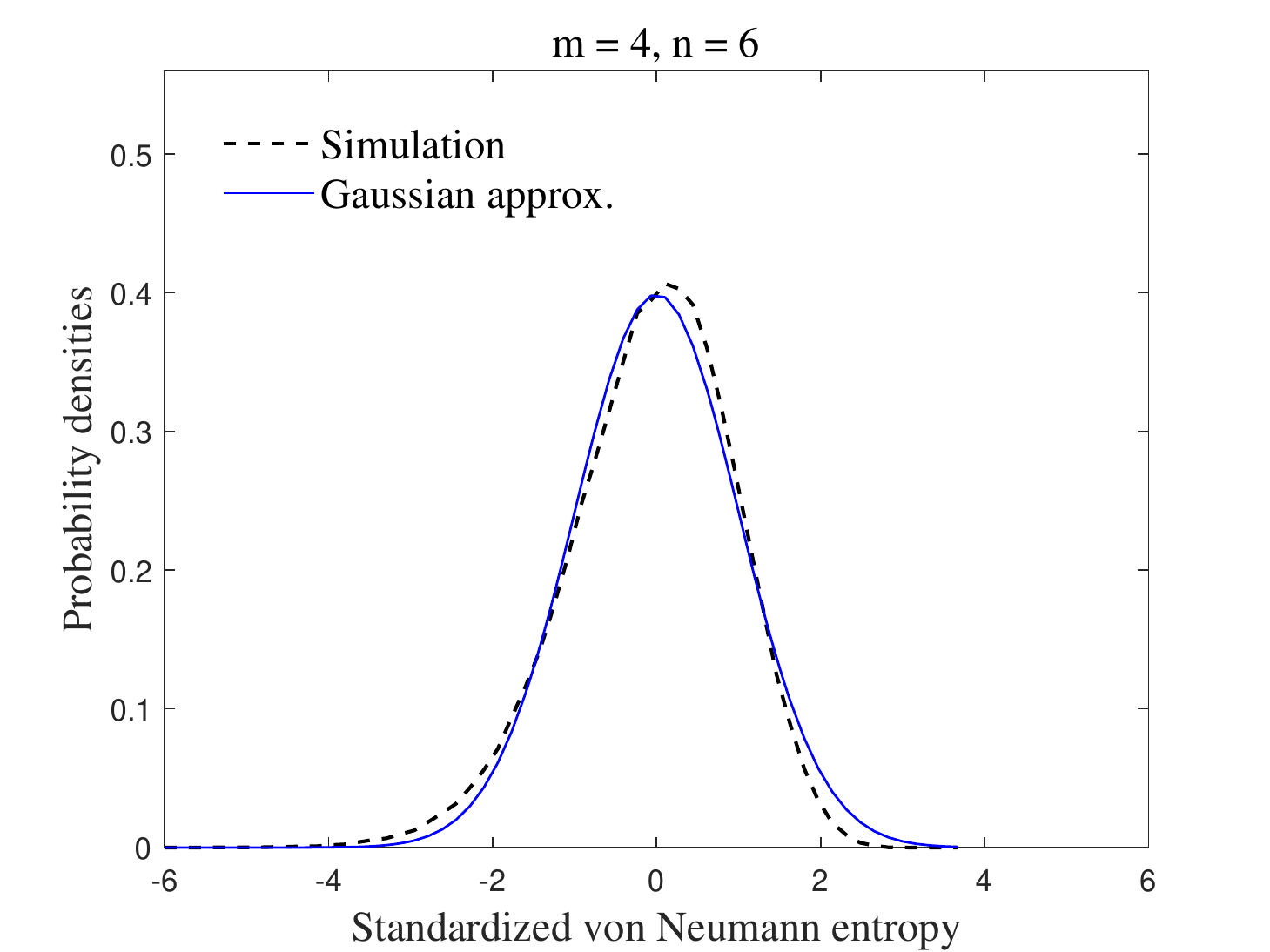}
\caption{Probability densities of standardized von Neumann entropy~(\ref{eq:X}) of subsystem dimensions $m=4$ and $n=6$: A comparison of the
simulated true distribution (dashed line in black) and the Gaussian approximation~(\ref{eq:Gau}) (solid line in blue).}
\label{fig:p1}
\end{figure}
In Fig.~\ref{fig:p1}, we compare the simulated true distribution of standardized von Neumann entropy~(\ref{eq:X}) to the Gaussian approximation~(\ref{eq:Gau}), where the dimensions of the subsystems are $m=4$ and $n=6$. As opposed to the Gaussian distribution, we see from Fig.~\ref{fig:p1} that the true distribution of von Neumann entropy is non-symmetric, which appears to be left-skewed (a negative skewness). With the knowledge of higher order moments, the Gaussian approximation~(\ref{eq:Gau}) can be systematically improved to provide more accurate approximations to finite-size systems. On the other hand, motivated by the case of Hilbert-Schmidt measure~\cite{Wei20}, here we also conjecture that the first two moments Eqs.~(\ref{eq:vNm}) and~(\ref{eq:vNv}) are sufficient to fully describe the distribution of von Neumann entropy as the dimensions of subsystems become large. Formally, in the limit
\begin{equation}\label{eq:lim}
m\to\infty,~~~~n\to\infty,~~~~\frac{m}{n}=c\in(0,1],
\end{equation}
\begin{figure}[!h]
\centering
\includegraphics[width=0.7\linewidth]{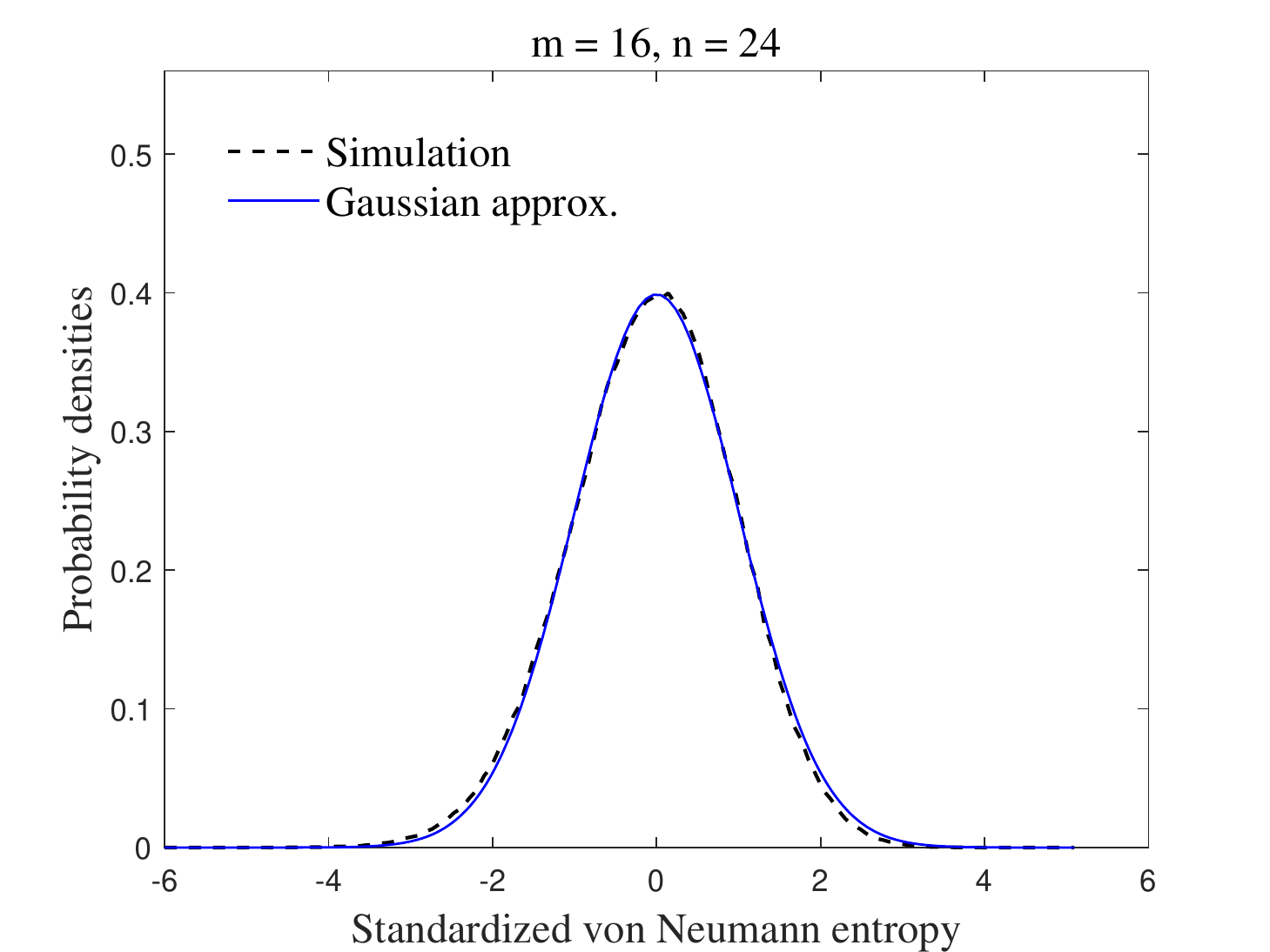}
\caption{Probability densities of standardized von Neumann entropy~(\ref{eq:X}) of subsystem dimensions $m=16$ and $n=24$: A numerical support to the conjectured Gaussian limit. The dashed line in black and the solid line in blue represents the simulated true distribution and the standard Gaussian distribution~(\ref{eq:Gau}), respectively.}
\label{fig:p2}
\end{figure}
we conjecture that the standardized von Neumann entropy~(\ref{eq:X}) converges in distribution to a Gaussian random variable with zero mean and unit variance. Note that the high-dimensional asymptotic regime~(\ref{eq:lim}) is different from the classical asymptotic regime~\cite{Page93}, where the dimension $m$ is fixed as $n$ goes to infinity. One way to prove the above conjecture is to show that all the higher order (beyond the first two) moments of the random variable~(\ref{eq:X}) vanish in the limit~(\ref{eq:lim}). A numerical evidence to support the conjecture is provided in Fig.~\ref{fig:p2}, where we simultaneously increase the subsystem dimensions to $m=16$ and $n=24$ with their ratio $c=m/n=2/3$ kept the same as in Fig.~\ref{fig:p1}. Comparing Fig.~\ref{fig:p1} with Fig.~\ref{fig:p2}, it is seen that the distribution of von Neumann entropy approaches rather rapidly to the conjectured limiting Gaussian distribution.

The rest of the paper is organized as follows. In Sec.~\ref{sec:vNv}, we derive the main result~(\ref{eq:vNv}) on the exact variance of von Neumann entropy over the Bures-Hall measure. Specifically, in Sec.~\ref{sec:rela} we relate the computation of the variance to that over a more convenient ensemble with no delta function constraint. Calculating the corresponding variance boils down in Sec.~\ref{sec:uvar} to computing four integrals over the correlation functions of the unconstraint ensemble. In Sec.~\ref{sec:IABCD}, the four integrals are evaluated into terms involving polygamma functions by utilizing recent results on the unconstraint ensemble as well as some summation formulas of polygamma functions. We outline potential future works in Sec.~\ref{sec:con} after summarizing the main findings of the paper. The polygamma summation formulas utilized in Sec.~\ref{sec:IABCD} are listed and discussed in Appendix~\ref{app:1}.

\section{Variance Calculation}\label{sec:vNv}
\subsection{Variance relation}\label{sec:rela}
Finding moment relations is a rather standard calculation, see, e.g., Refs.~\cite{Sarkar19,Wei20a,Osipov10,Page93,Ruiz95,Wei17,Wei20}, that relates moment computation to that over an ensemble without the constraint $\delta\left(1-\sum_{i=1}^{m}\lambda_{i}\right)$. As will be seen, the unconstrained ensemble of the Bures-Hall measure~(\ref{eq:BH}) is~\cite{Sarkar19}
\begin{equation}\label{eq:BHu}
h\left(\bm{x}\right)=\frac{1}{C'}\prod_{1\leq i<j\leq m}\frac{\left(x_{i}-x_{j}\right)^{2}}{x_{i}+x_{j}}\prod_{i=1}^{m}x_{i}^{\alpha}\e^{-x_{i}},
\end{equation}
where $x_{i}\in[0,\infty)$, $i=1,\dots,m$, and the constant $C'$ depends on the constant~(\ref{eq:cBH}) as
\begin{equation}\label{eq:cBHu}
C'=C~\Gamma\left(d\right)
\end{equation}
with $d$ denoting
\begin{equation}\label{eq:d}
d=\frac{1}{2}m\left(m+2\alpha+1\right).
\end{equation}
Despite being only interested in the physically relevant $\alpha$ values in Eq.~(\ref{eq:aBH}), the results hereafter, in particular the expression~(\ref{eq:Tva}), are valid for any $\alpha>-1$ that the density~(\ref{eq:BHu}) is defined.

We first derive the density $g(\theta)$ of trace
\begin{equation}\label{eq:tr}
\theta=\sum_{i=1}^{m}x_{i},~~~~\theta\in[0,\infty),
\end{equation}
of the unconstraint ensemble~(\ref{eq:BHu}) as
\begin{eqnarray}
g(\theta)&=&\int_{\bm{x}}h(\bm{x})\delta\left(\theta-\sum_{i=1}^{m}x_{i}\right)\prod_{i=1}^{m}\dd x_{i}\\
&=&\frac{C}{C'}\e^{-\theta}\theta^{d-1}\int_{\bm{\lambda}}f(\bm{\lambda})\prod_{i=1}^{m}\dd\lambda_{i}\\
&=&\frac{1}{\Gamma\left(d\right)}\e^{-\theta}\theta^{d-1},\label{eq:g}
\end{eqnarray}
where we have employed the change of variables
\begin{equation}\label{eq:cv}
x_{i}=\theta\lambda_{i},~~~~i=1,\ldots,m.
\end{equation}
The above calculation implies that the density $h(\bm{x})$ is factored as~\footnote{A similar factorization in the Hilbert-Schmidt measure also exists~\cite{Page93}.}
\begin{equation}\label{eq:g2ft}
h(\bm{x})\prod_{i=1}^{m}\dd x_{i}=f(\bm{\lambda})g(\theta)\dd\theta\prod_{i=1}^{m}\dd\lambda_{i},
\end{equation}
which leads to the fact that $\theta$ is independent of each $\lambda_{i}$ (hence independent of $S$).

To exploit this independence in calculating the variance, we first write by the change of variables~(\ref{eq:cv}) that
\begin{equation}\label{eq:ST}
S^{2}=\theta^{-2}T^{2}+2S\ln\theta-\ln^{2}\theta,
\end{equation}
where
\begin{equation}\label{eq:TvN}
T=\sum_{i=1}^{m}x_{i}\ln x_{i}
\end{equation}
defines the induced von Neumann entropy over the unconstrained ensemble~(\ref{eq:BHu}). The second moment relation can now be found, by multiplying an appropriate constant (cf.~Eq.~(\ref{eq:g}))
\begin{equation}
1=\int_{0}^{\infty}\frac{1}{\Gamma\left(d+2\right)}\e^{-\theta}\theta^{d+1}\dd\theta
\end{equation}
as
\begin{eqnarray}
\mathbb{E}_{f}\!\left[S^{2}\right]&=&\int_{0}^{\infty}\!\!\int_{\bm{\lambda}}\frac{\e^{-\theta}\theta^{d+1}}{\Gamma\left(d+2\right)}S^{2}f(\bm{\lambda})\dd\theta\prod_{i=1}^{m}\dd\lambda_{i}\\
&=&\frac{\Gamma(d)}{\Gamma(d+2)}\mathbb{E}_{h}\!\left[T^{2}\right]+2\mathbb{E}_{f}\!\left[S\right]\mathbb{E}_{g}\!\left[\ln\theta\right]-\mathbb{E}_{g}\!\left[\ln^{2}\theta\right],\nonumber\\
&=&\frac{1}{d(d+1)}\mathbb{E}_{h}\!\left[T^{2}\right]+2\psi_{0}(d+2)\mathbb{E}_{f}\!\left[S\right]-\psi_{0}^{2}(d+2)-\psi_{1}(d+2),\label{eq:S2T}
\end{eqnarray}
where we have used the results Eqs.~(\ref{eq:g2ft}),~(\ref{eq:ST}) and the identities (valid for $\Re(a)>0$)
\begin{subequations}
\begin{eqnarray}
\int_{0}^{\infty}\!\!\e^{-\theta}\theta^{a-1}\ln{\theta}\dd\theta&=&\Gamma(a)\psi_{0}(a),\\
\int_{0}^{\infty}\!\!\e^{-\theta}\theta^{a-1}\ln^{2}{\theta}\dd\theta&=&\Gamma(a)\left(\psi_{0}^{2}(a)+\psi_{1}(a)\right).
\end{eqnarray}
\end{subequations}
By the known mean formulas~(\ref{eq:vNm}) and~\cite{Wei20a}
\begin{equation}\label{eq:Tm}
\mathbb{E}_{h}\!\left[T\right]=\frac{m(m+2\alpha+1)}{2}\psi_{0}(m+\alpha+1),
\end{equation}
the derived moment relation~(\ref{eq:S2T}) translates showing the claimed variance formula~(\ref{eq:vNv}) to proving an induced variance formula
\begin{flalign}\label{eq:Tv}
\mathbb{V}\!_{h}\!\left[T\right]=m(2n-m)\Bigg(\psi_{0}\left(n+\frac{1}{2}\right)+\frac{1}{2}\psi_{0}^{2}\left(n+\frac{1}{2}\right)+\frac{4n^{2}+2mn-m^{2}+1}{8n}\psi_{1}\left(n+\frac{1}{2}\right)\!\Bigg),&&
\end{flalign}
where we have also used the fact $\mathbb{V}\!\left[X\right]=\mathbb{E}\!\left[X^{2}\right]-\mathbb{E}^{2}\!\left[X\right]$ and the identities
\begin{subequations}\label{eq:pgre}
\begin{eqnarray}
\psi_{0}(l+n)&=&\psi_{0}(l)+\sum_{k=0}^{n-1}\frac{1}{l+k},\label{eq:dgre}\\
\psi_{1}(l+n)&=&\psi_{1}(l)-\sum_{k=0}^{n-1}\frac{1}{(l+k)^2}.\label{eq:tgre}
\end{eqnarray}
\end{subequations}

\subsection{Variance of unconstraint ensemble}\label{sec:uvar}
Calculating $\mathbb{E}_{h}\!\left[T^{2}\right]$ requires one and two arbitrary eigenvalue densities, denoted respectively by $h_{1}(x)$ and $h_{2}(x,y)$, of the unconstraint Bures-Hall ensemble~(\ref{eq:BHu}) as
\begin{equation}\label{eq:i2m}
\mathbb{E}_{h}\!\left[T^{2}\right]=m\int_{0}^{\infty}\!\!x^{2}\ln^{2}x~h_{1}(x)\dd x+m(m-1)\int_{0}^{\infty}\!\!\int_{0}^{\infty}\!\!xy\ln{x}\ln{y}~h_{2}\left(x,y\right)\dd x\dd y.
\end{equation}
In general, the density of $k$ arbitrary eigenvalues ($k$-point correlation function) of the ensemble~(\ref{eq:BHu}) is described by a Pfaffian point process of a $2k\times 2k$ anti-symmetric matrix~\cite{FK16}. The corresponding correlation kernels are written in terms of these of the Cauchy-Laguerre biorthogonal ensemble~\cite{Bertola14}, which is a determinantal point process. In particular, the needed eigenvalue densities in Eq.~(\ref{eq:i2m}) are written as~\cite{FK16,Bertola14}
\begin{eqnarray}
h_{1}(x)&=&\frac{1}{2m}\left(K_{01}(x,x)+K_{10}(x,x)\right),\label{eq:h1} \\
h_{2}(x,y)&=&\frac{1}{4m(m-1)}(\left(K_{01}(x,x)+K_{10}(x,x)\right)\left(K_{01}(y,y)+K_{10}(y,y)\right)-2K_{01}(x,y)K_{01}(y,x) \nonumber\\
&&-2K_{10}(x,y)K_{10}(y,x)-2K_{00}(x,y)K_{11}(x,y)-2K_{00}(y,x)K_{11}(y,x)),\label{eq:h2}
\end{eqnarray}
where the correlation kernels are
\begin{subequations}\label{eq:ker}
\begin{eqnarray}
K_{00}(x,y)&=&\sum_{k=0}^{m-1}p_{k}(x)q_{k}(y), \label{eq:K00} \\
K_{01}(x,y)&=&-x^{2\alpha+1}y^{-\alpha-1}e^{-y}\sum_{k=0}^{m-1}p_{k}(x)Q_{k}(-y), \label{eq:K01} \\
K_{10}(x,y)&=&-x^{-\alpha}y^{2\alpha+1}e^{-x}\sum_{k=0}^{m-1}P_{k}(-x)q_{k}(y), \label{eq:K10} \\
K_{11}(x,y)&=&x^{\alpha}y^{\alpha+1}e^{-x-y}\sum_{k=0}^{m-1}P_{k}(-y)Q_{k}(-x)-w(x,y) \label{eq:K11}
\end{eqnarray}
\end{subequations}
with the weight function $w(x,y)$ of the biorthogonal polynomials $p_{k}(x)$, $q_{l}(y)$,
\begin{equation}\label{eq:oc}
\int_{0}^{\infty}\!\!\int_{0}^{\infty}p_{k}(x)q_{l}(y)w(x,y)\dd x\dd y=\delta_{kl}
\end{equation}
given by
\begin{equation}\label{eq:w}
w(x,y)=\frac{x^{\alpha}y^{\alpha+1}e^{-x-y}}{x+y}.
\end{equation}
The functions in Eq.~(\ref{eq:ker}) are further related by~\cite{FK16,Bertola14}
\begin{subequations}\label{eq:PQ}
\begin{eqnarray}
P_{k}(x)&=&\int_{0}^{\infty}\frac{v^{\alpha}\e^{-v}}{x-v}p_{k}(v)\dd v, \\
Q_{k}(y)&=&\int_{0}^{\infty}\frac{w^{\alpha+1}\e^{-w}}{y-w}q_{k}(w)\dd w,
\end{eqnarray}
\end{subequations}
which, together with the orthogonality condition~(\ref{eq:oc}), can be employed to verify that the functions~(\ref{eq:h1}) and~(\ref{eq:h2}) are indeed probability density functions. Moreover, these function are expressed explicitly via Meijer G-functions as~\cite{FK16,Bertola14}
\begin{subequations}\label{eq:kerM}
\begin{eqnarray*}
p_{j}(x)&=&\sqrt{2}(-1)^{j}G_{2,3}^{1,1}\left(\!\begin{array}{c}-2\alpha-1-j;j+1\\0;-\alpha,-2\alpha-1\end{array}\bigg|x\Big.\right),\\
q_{j}(x)&=&\sqrt{2}(-1)^{j}(j+\alpha+1)G_{2,3}^{1,1}\left(\!\begin{array}{c}-2\alpha-1-j;j+1\\0;-\alpha-1,-2\alpha-1\end{array}\bigg|x\Big.\right),\\
P_{j}(x)&=&\sqrt{2}(-1)^{j+1}\e^{-x}G_{2,3}^{2,1}\left(\!\begin{array}{c}-\alpha-j-1;\alpha+j+1\\0,\alpha;-\alpha-1\end{array}\bigg|-\!x\Big.\right),\\
Q_{j}(x)&=&\sqrt{2}(-1)^{j+1}(j+\alpha+1)\e^{-x}G_{2,3}^{2,1}\left(\!\begin{array}{c}-\alpha-j;\alpha+j+2\\0,\alpha+1;-\alpha\end{array}\bigg|-\!x\Big.\right),
\end{eqnarray*}
\end{subequations}
where the Meijer G-function is defined by the contour integral~\cite{Prudnikov86}
\begin{flalign}\label{eq:MG}
G_{p,q}^{m,n}\left(\begin{array}{c} a_{1},\ldots,a_{n}; a_{n+1},\ldots,a_{p} \\ b_{1},\ldots,b_{m}; b_{m+1},\ldots,b_{q} \end{array}\bigg|x\Big.\right)=\frac{1}{2\pi\imath}\int_{\mathcal{L}}{\frac{\prod_{j=1}^m\Gamma\left(b_j+s\right)\prod_{j=1}^n\Gamma\left(1-a_j-s\right)x^{-s}}{\prod_{j=n+1}^p \Gamma\left(a_{j}+s\right)\prod_{j=m+1}^q\Gamma\left(1-b_j-s\right)}}\dd s &&
\end{flalign}
with the contour $\mathcal{L}$ separating the poles of $\Gamma\left(1-a_j-s\right)$ from the poles of $\Gamma\left(b_j+s\right)$. Beside the summation form~(\ref{eq:ker}), the kernels also admit integral representation~\cite{FK16,Bertola14}
\begin{subequations}\label{eq:kerI}
\begin{eqnarray}
K_{00}(x,y)&=&\int_{0}^{1}t^{2\alpha+1}H_{\alpha}(tx)H_{\alpha+1}(ty)\dd t, \label{eq:K00I} \\
K_{01}(x,y)&=&x^{2\alpha+1}\int_{0}^{1}t^{2\alpha+1}H_{\alpha}(tx)G_{\alpha+1}(ty)\dd t, \label{eq:K01I} \\
K_{10}(x,y)&=&y^{2\alpha+1}\int_{0}^{1}t^{2\alpha+1}G_{\alpha}(tx)H_{\alpha+1}(ty)\dd t, \label{eq:K10I} \\
K_{11}(x,y)&=&(xy)^{2\alpha+1}\int_{0}^{1}t^{2\alpha+1}G_{\alpha+1}(tx)G_{\alpha}(ty)\dd t-\frac{x^{\alpha}y^{\alpha+1}}{x+y}, \label{eq:K11I}
\end{eqnarray}
\end{subequations}
where we denote
\begin{subequations}
\begin{eqnarray}
H_{q}(x)&=&G_{2,3}^{1,1}\left(\!\begin{array}{c}-m-2\alpha-1;m\\0;-q,-2\alpha-1\end{array}\bigg|x\Big.\right),\\
G_{q}(x)&=&G_{2,3}^{2,1}\left(\!\begin{array}{c}-m-2\alpha-1;m\\0,-q;-2\alpha-1\end{array}\bigg|x\Big.\right).
\end{eqnarray}
\end{subequations}

Finally, inserting Eqs.~(\ref{eq:h1}) and~(\ref{eq:h2}) into Eq.~(\ref{eq:i2m}), the induced variance is represented as
\begin{equation}\label{eq:Tvi}
\mathbb{V}\!_{h}\!\left[T\right]=\frac{1}{2}\left(I_{A}-I_{B}-I_{C}+2I_{D}\right),
\end{equation}
where
\begin{eqnarray}
\!\!\!\!\!\!\!\!I_{A}&=&\int_{0}^{\infty}x^{2}\ln^{2}x\left(K_{01}(x,x)+K_{10}(x,x)\right)\dd x, \label{eq:IA} \\
\!\!\!\!\!\!\!\!I_{B}&=&\int_{0}^{\infty}\!\!\int_{0}^{\infty}xy\ln x\ln y~K_{01}(x,y)K_{01}(y,x)\dd x\dd y, \label{eq:IB} \\
\!\!\!\!\!\!\!\!I_{C}&=&\int_{0}^{\infty}\!\!\int_{0}^{\infty}xy\ln x\ln y~K_{10}(x,y)K_{10}(y,x)\dd x\dd y, \label{eq:IC} \\
\!\!\!\!\!\!\!\!I_{D}&=&\int_{0}^{\infty}\!\!\int_{0}^{\infty}xy\ln x\ln y~K_{00}(x,y)K_{11}(x,y)\dd x\dd y, \label{eq:ID}
\end{eqnarray}
and we have used the fact (cf.~Eq.~(\ref{eq:Tm}))
\begin{equation}
\int_{0}^{\infty}x\ln x\left(K_{01}(x,x)+K_{10}(x,x)\right)\dd x=2\mathbb{E}_{h}\!\left[T\right].
\end{equation}
To show Eq.~(\ref{eq:Tv}), the remaining task is to compute the four integrals~(\ref{eq:IA})--(\ref{eq:ID}).

\subsection{Computing the integrals $I_{A}$, $I_{B}$, $I_{C}$, and $I_{D}$}\label{sec:IABCD}
\subsubsection{Computation of $I_{A}$}
The evaluation of $I_{A}$ follows a similar procedure as in Ref.~\cite[Sec.~2.2]{Wei20a}. The key is to compute the integral
\begin{equation}\label{eq:Aq}
A_{q}(t)=\int_{0}^{\infty}x^{\beta}(tx)^{2\alpha+1}H_{2\alpha+1-q}(tx)G_{q}(tx)\dd x
\end{equation}
and its derivatives with respect to $\beta$ for $q=\alpha, \alpha+1$. This integral has been obtained in Ref.~\cite{Wei20a} by using the Mellin transform of Meijer G-function~\cite{Prudnikov86}
\begin{flalign}\label{eq:iMG}
\int_{0}^{\infty}x^{s-1}G_{p,q}^{m,n}\left(\begin{array}{c} a_{1},\ldots,a_{n}; a_{n+1},\ldots,a_{p} \\ b_{1},\ldots,b_{m}; b_{m+1},\ldots,b_{q} \end{array}\bigg|\eta x\Big.\right)\dd x=\frac{\eta^{-s}\prod_{j=1}^m\Gamma\left(b_j+s\right)\prod_{j=1}^{n}\Gamma\left(1-a_j-s\right)}{\prod_{j=n+1}^{p}\Gamma\left(a_{j}+s\right)\prod_{j=m+1}^q\Gamma\left(1-b_j-s\right)}&&
\end{flalign}
and the fact that the Meijer G-function $G_{2,3}^{1,1}$ of a negative parameter $a_{i}$ ($i\leq n$) is a terminating hypergeometric function~\cite{Bertola14,Wei20a}~\footnote{This fact is also the starting point in computing $I_{B}$, $I_{C}$, and $I_{D}$.} as
\begin{equation}\label{eq:Ibs0}
A_{q}(t)=t^{-\beta-1}A_{q},
\end{equation}
where
\begin{flalign}\label{eq:Ibs}
A_{q}=\sum_{k=0}^{m-1}\frac{(-1)^{k+m}\Gamma(k+2\alpha+m+2)\Gamma(k+\beta+1)}{\Gamma(k+2\alpha+2)\Gamma(k+2\alpha+2-q)\Gamma(m-k)k!}\frac{\Gamma(k+\beta+2\alpha+2)\Gamma(k+\beta+2\alpha+2-q)}{\Gamma(k+\beta+2\alpha+m+2)\Gamma(k+\beta-m+1)}.&&
\end{flalign}
Using a different representation of Eqs.~(\ref{eq:K01I}) and~(\ref{eq:K10I}) obtained in Ref.~\cite{Wei20a},
\begin{subequations}
\begin{eqnarray*}
K_{01}(x,y)&=&-x^{2\alpha+1}\int_{1}^{\infty}t^{2\alpha+1}H_{\alpha}(tx)G_{\alpha+1}(ty)\dd t,\\
K_{10}(x,y)&=&-y^{2\alpha+1}\int_{1}^{\infty}t^{2\alpha+1}G_{\alpha}(tx)H_{\alpha+1}(ty)\dd t, \\
\end{eqnarray*}
\end{subequations}
and changing the order of integrations, $I_{A}$ is calculated as
\begin{equation}\label{eq:IAi}
I_{A}=-\frac{1}{4}\left(A_{\alpha}+A_{\alpha+1}\right)+\frac{1}{2}\left(H_{\alpha}^{(1)}+H_{\alpha+1}^{(1)}\right)-\frac{1}{2}\left(H_{\alpha}^{(2)}+H_{\alpha+1}^{(2)}\right),
\end{equation}
where we denote
\begin{equation}\label{eq:H12}
H_{q}^{(1)}=\frac{\dd}{\dd\beta}A_{q}^{(\beta)}\Big|_{\beta=2},~~~~~~H_{q}^{(2)}=\frac{\dd}{\dd\beta^{2}}A_{q}^{(\beta)}\Big|_{\beta=2},
\end{equation}
and the integrals over $t$ have been evaluated first by the fact
\begin{equation*}
\int_{1}^{\infty}\frac{1}{t^{3}}\dd t=\frac{1}{2},~~~~~~\int_{1}^{\infty}\frac{\ln t}{t^{3}}\dd t=\frac{1}{4},~~~~~~\int_{1}^{\infty}\frac{\ln^{2} t}{t^{3}}\dd t=\frac{1}{4}.
\end{equation*}
The first term $A_{\alpha}+A_{\alpha+1}$ in Eq.~(\ref{eq:IAi}) has been obtained in Ref.~\cite[Eq.~(50)]{Wei20a}. By resolving indeterminacy in the limit $\epsilon\to0$,
\begin{subequations}\label{eq:pgna}
\begin{eqnarray}
\Gamma(-l+\epsilon)&=&\frac{(-1)^{l}}{l!\epsilon}\left(1+\psi_{0}(l+1)\epsilon+o\left(\epsilon^2\right)\right),\label{eq:pgna1}\\
\psi_{0}(-l+\epsilon)&=&-\frac{1}{\epsilon}\left(1-\psi_{0}(l+1)\epsilon+o\left(\epsilon^2\right)\right),\label{eq:pgna2}\\
\psi_{1}(-l+\epsilon)&=&\frac{1}{\epsilon^2}\left(1+o\left(\epsilon^2\right)\right),\label{eq:pgna3}
\end{eqnarray}
\end{subequations}
the terms~(\ref{eq:H12}) are evaluated into finite sums involving polygamma functions. Computing these summations by the identities in Appendix~\ref{app:1}, we obtain $I_{A}$ as shown in Eq.~(\ref{eq:IAf}), where the list of coefficients can be found in Table~\ref{t:IA} of Appendix~\ref{app:2}. Note that as a result of employing the semi-closed-form identities~(\ref{eq:A6})--(\ref{eq:A8}), the obtained $I_{A}$ expression~(\ref{eq:IAf}) still contains five summations that may not be further simplified. These unsimplifiable sums eventually cancel with the ones in $I_{B}$ and $I_{C}$ as will be seen. Similar phenomena have also been observed in the higher order moment computations over the Hilbert-Schmidt measure~\cite{Wei17,Wei20}.
\begin{eqnarray}\label{eq:IAf}
I_{A}&=&\frac{1}{36\alpha(m+\alpha)(m+2\alpha)(2m+2\alpha+1)^3}\Bigg(\!\!-2a_{0}\Bigg(\sum_{k=1}^{m}\frac{\psi_{0}(k+\alpha)}{k}+\sum_{k=1}^{m}\frac{\psi_{0}(k+2\alpha)}{k}\nonumber\\
&&-\sum_{k=1}^{m}\frac{\psi_{0}(k+m+2\alpha)}{k}+\sum_{k=1}^{m}\frac{\psi_{0}(k+m+2\alpha)}{k+\alpha}+\sum_{k=1}^{m}\frac{\psi_{0}(k+m+2\alpha)}{k+2\alpha}\Bigg)+a_{1}+a_{2}(\psi_{0}(1)\nonumber\\
&&-\psi_{0}(m+1))+a_{3}\psi_{0}(\alpha+1)+a_{4}\psi_{0}(2\alpha+1)+a_{5}\psi_{0}(m+\alpha+1)+a_{6}\psi_{0}(m+2\alpha+1)\nonumber\\
&&+a_{7}\psi_{0}(2m+2\alpha+1)+a_{0}\Big(\!-2\psi_{0}(1)\psi_{0}(\alpha+1)-2\psi_{0}(1)\psi_{0}(2\alpha+1)+2\psi_{0}(1)\nonumber\\
&&\times\psi_{0}(m+2\alpha+1)+\psi_{0}^{2}(\alpha+1)+\psi_{0}^{2}(2\alpha+1)+2\psi_{0}(\alpha+1)\psi_{0}(m+1)-2\psi_{0}(\alpha+1)\nonumber\\
&&\times\psi_{0}(m+\alpha+1)-2\psi_{0}(\alpha+1)\psi_{0}(m+2\alpha+1)+2\psi_{0}(2\alpha+1)\psi_{0}(m+1)-4\psi_{0}(2\alpha+1)\nonumber\\
&&\times\psi_{0}(m+2\alpha+1)-2\psi_{0}(m+1)\psi_{0}(m+2\alpha+1)-2\psi_{0}(m+\alpha+1)\psi_{0}(m+2\alpha+1)\nonumber\\
&&+4\psi_{0}(m+\alpha+1)\psi_{0}(2m+2\alpha+1)-\psi_{0}^{2}(m+2\alpha+1)+8\psi_{0}(m+2\alpha+1)\psi_{0}(2m\nonumber\\
&&+2\alpha+1)-4\psi_{0}^{2}(2m+2\alpha+1)-\psi_{1}(\alpha+1)-\psi_{1}(2\alpha+1)+\psi_{1}(m+2\alpha+1)\Big)\!\Bigg).
\end{eqnarray}

\subsubsection{Computation of $I_{B}$ and $I_{C}$}
The steps in calculating $I_{B}$ and $I_{C}$ are identical. The starting point is the integral form of the kernels~(\ref{eq:K01I}),~(\ref{eq:K10I}) as well as finite sum representation~\cite{Bertola14,Wei20a} of the Meijer G-functions $G_{2,3}^{1,1}$. Instead of changing the order of summations as in $I_{A}$, here we directly evaluate the integrals over $t$ by the identity~\cite{Prudnikov86}
\begin{eqnarray}
&&\int_{0}^{1}\!x^{a-1}G_{p,q}^{m,n}\left(\begin{array}{c} a_{1},\ldots,a_{n}; a_{n+1},\ldots,a_{p} \\ b_{1},\ldots,b_{m}; b_{m+1},\ldots,b_{q} \end{array}\bigg|\eta x\Big.\right)\dd x \nonumber\\
&=&G_{p+1,q+1}^{m,n+1}\left(\begin{array}{c}1-a,a_{1},\ldots,a_{n};a_{n+1},\ldots,a_{p}\\b_{1},\ldots,b_{m};b_{m+1},\ldots,b_{q},-a\end{array}\bigg|\eta\Big.\right).
\end{eqnarray}
This leads $I_{B}$ to
\begin{equation}
I_{B}=\sum_{j,k=0}^{m-1}f_{j,k}f_{k,j},
\end{equation}
where we denote
\begin{eqnarray*}
&&f_{j,k}=\frac{(-1)^{j}\Gamma(m+2\alpha+j+2)}{\Gamma(j+1)\Gamma(\alpha+j+1)\Gamma(2\alpha+j+2)\Gamma(m-j)}\nonumber\\
&&\times\int_{0}^{\infty}\!x\ln x~G_{3,4}^{2,2}\left(\!\begin{array}{c}j-k,j-m;m+2\alpha+j+1\\2\alpha+j+1,\alpha+j;j,j-k-1\end{array}\bigg|x\Big.\right)\dd x.
\end{eqnarray*}
The above integral can be similarly evaluated as in Eq.~(\ref{eq:Aq}) by first utilizing Eq.~(\ref{eq:iMG}) before taking the derivative with respect to $\beta$. We then set $\beta=1$ and resolve the resulting indeterminacy by Eq.~(\ref{eq:pgna}), $I_{B}$ becomes a double sum involving polygamma functions. The summations are evaluated with the help of the identities in Appendix~\ref{app:1}, which completes the calculation of $I_{B}$. Since $I_{C}$ is computed to the same form as $I_{B}$, for convenience we provide the corresponding result of $I_{B}+I_{C}$ as shown in Eq.~(\ref{eq:IBCf}), where the coefficients can be found in Table~\ref{t:IBC}.

\begin{eqnarray}\label{eq:IBCf}
I_{B}+I_{C}&=&\frac{1}{36\alpha(m+\alpha)(m+\alpha+1)(m+2\alpha)(2m+2\alpha+1)^4}\Bigg(2\left(b_{0}+c_{0}\right)\Bigg(\sum_{k=1}^{m}\frac{\psi_{0}(k+\alpha)}{k}\nonumber\\
&&+\sum_{k=1}^{m}\frac{\psi_{0}(k+2\alpha)}{k}-\sum_{k=1}^{m}\frac{\psi_{0}(k+m+2\alpha)}{k}+\sum_{k=1}^{m}\frac{\psi_{0}(k+m+2\alpha)}{k+\alpha}\nonumber\\
&&+\sum_{k=1}^{m}\frac{\psi_{0}(k+m+2\alpha)}{k+2\alpha}\Bigg)+b_{1}+c_{1}+\left(b_{2}+c_{2}\right)\left(\psi_{0}(1)-\psi_{0}(m+1)\right)+(b_{3}+c_{3})\nonumber\\
&&\times\psi_{0}(\alpha+1)+(b_{4}+c_{4})\psi_{0}(2\alpha+1)+(b_{5}+c_{5})\psi_{0}(m+\alpha+1)+(b_{6}+c_{6})\nonumber\\
&&\times\psi_{0}(m+2\alpha+1)+(b_{7}+c_{7})\psi_{0}(2m+2\alpha+1)+(b_{0}+c_{0})\Big(2\psi_{0}(1)\psi_{0}(\alpha+1)\nonumber\\
&&+2\psi_{0}(1)\psi_{0}(2\alpha+1)-2\psi_{0}(1)\psi_{0}(m+2\alpha+1)-\psi_{0}^{2}(\alpha+1)-\psi_{0}^{2}(2\alpha+1)\nonumber\\
&&-2\psi_{0}(\alpha+1)\psi_{0}(m+1)+2\psi_{0}(\alpha+1)\psi_{0}(m+\alpha+1)+2\psi_{0}(\alpha+1)\psi_{0}(m+2\alpha+1)\nonumber\\
&&-2\psi_{0}(2\alpha+1)\psi_{0}(m+1)+4\psi_{0}(2\alpha+1)\psi_{0}(m+2\alpha+1)+2\psi_{0}(m+1)\nonumber\\
&&\times\psi_{0}(m+2\alpha+1)+\psi_{1}(\alpha+1)+\psi_{1}(2\alpha+1)-\psi_{1}(m+\alpha+1)-3\psi_{1}(m+2\alpha+1)\nonumber\\
&&+2\psi_{1}(2m+2\alpha+1)\Big)+(b_{8}+c_{8})\psi_{0}^{2}(m+\alpha+1)+(b_{9}+c_{9})\psi_{0}(m+\alpha+1)\nonumber\\
&&\times\psi_{0}(m+2\alpha+1)+(b_{10}+c_{10})\Big(\psi_{0}(m+\alpha+1)\psi_{0}(2m+2\alpha+1)+2\psi_{0}(m+2\alpha\nonumber\\
&&+1)\psi_{0}(2m+2\alpha+1)-\psi_{0}^{2}(2m+2\alpha+1)\Big)+(b_{11}+c_{11})\psi_{0}^{2}(m+2\alpha+1)\Bigg).
\end{eqnarray}

\subsubsection{Computation of $I_{D}$}
We define the integral
\begin{equation}\label{eq:IDg}
D\left(\beta_{1},\beta_{2}\right)=\int_{0}^{\infty}\!\!\int_{0}^{\infty}x^{\beta_{1}}y^{\beta_{2}}~K_{00}(x,y)K_{11}(x,y)\dd x\dd y
\end{equation}
so that the desired $I_{D}$ integral~(\ref{eq:ID}) can be obtained as
\begin{equation}\label{eq:IDr}
I_{D}=\frac{\partial^{2}}{\partial\beta_{1}\partial\beta_{2}}D\left(\beta_{1},\beta_{2}\right)\Big|_{\beta_{1}=1,\beta_{2}=1}.
\end{equation}
To compute the integral~(\ref{eq:IDg}), one uses the summation form of the kernels~(\ref{eq:ker}) instead of the integral representation~(\ref{eq:kerI}). The corresponding integrals over $x$ and $y$ can then be separately evaluated by the formula~(\ref{eq:iMG}) and explicit expressions of the polynomials $p_{k}(x)$ and $q_{k}(y)$ in Ref.~\cite{Bertola14}. Now taking the partial derivatives~(\ref{eq:IDr}) gives
\begin{equation}
I_{D}=I_{D1}-I_{D2},
\end{equation}
where
\begin{equation}\label{eq:ID1}
I_{D1}=\lim_{\beta\to1}\sum_{j=0}^{m-1}\sum_{k=0}^{m-1}\sum_{i=0}^{j}\sum_{s=0}^j(j+\alpha+1)(k+\alpha+1)~g_{\alpha,i}~g_{\alpha+1,s},
\end{equation}
\begin{eqnarray}\label{eq:ID2}
I_{D2}&=&\sum_{j=0}^{m-1}\sum_{i=0}^{j}\sum_{s=0}^{j}\frac{2(i+\alpha+1)(s+\alpha+2)h_{i}h_{s}}{(j+\alpha+1)^{-1}(i+s+2\alpha+4)}\bigg(\!\psi_{0}(i+\alpha+2)\psi_{0}(s+\alpha+3)\nonumber\\
&&+\frac{2}{(i+s+2\alpha+4)^2}-\frac{\psi_{0}(i+\alpha+2)+\psi_{0}(s+\alpha+3)}{i+s+2\alpha+4}\bigg)
\end{eqnarray}
with the shorthand notations
\begin{eqnarray}
h_{r}&=&\frac{(-1)^{r}\Gamma(r+j+2\alpha+2)}{\Gamma(r+1)\Gamma(j-r+1)\Gamma(r+2\alpha+2)},\nonumber\\
g_{p,r}&=&\frac{2h_{r}\Gamma(r+\beta+1)\Gamma(p+r+\beta+1)\Gamma(r+2\alpha+\beta+2)}{\Gamma(r-k+\beta+1)\Gamma(r+k+2\alpha+\beta+3)\Gamma(p+r+1)}(\psi_{0}(p+r+\beta+1)\nonumber\\
&&+\psi_{0}(r+2\alpha+\beta+2)+\psi_{0}(r+\beta+1)-\psi_{0}(r-k+\beta+1)-\psi_{0}(r+k+2\alpha+\beta+3)). \nonumber
\end{eqnarray}
For the $I_{D1}$ sums~(\ref{eq:ID1}), the summation over $j$ is evaluated first by the identity~\cite[Lemma~4.1]{Bertola14}
\begin{flalign}\label{eq:L41}
\sum_{j=i}^{m-1}(j+\alpha+1)\frac{\Gamma(j+i+2\alpha+2)\Gamma(j+s+2\alpha+2)}{\Gamma(j-i+1)\Gamma(j-s+1)}=\frac{\Gamma(i+m+2\alpha+2)\Gamma(s+m+2\alpha+2)}{2(i+s+2\alpha+2)\Gamma(m-i)\Gamma(m-s)}.&&
\end{flalign}
After determining the limits when $\beta\to1$, the summation over $k$ is evaluated next by the identity
\begin{flalign}\label{eq:3t2}
\sum_{k=0}^{m}(k+\alpha+1)\frac{\left(\Gamma(s-k+2)\Gamma(k+s+2\alpha+4)\right)^{-1}}{\Gamma(i-k+2)\Gamma(k+i+2\alpha+4)}=\frac{\left(\Gamma(i+2\alpha+3)\Gamma(s+2\alpha+3)\right)^{-1}}{2\Gamma(i+2)\Gamma(s+2)(i+s+2\alpha+4)}&&
\end{flalign}
as well as three additional identities obtained by taking derivatives of Eq.~(\ref{eq:3t2}) with respect to $i$, $s$, and both $i$ and $s$. Now the $I_{D1}$ quadruple sum~(\ref{eq:ID1}) reduces to double sums in $i$ and $s$. Similarly, for the $I_{D2}$ sums~(\ref{eq:ID2}) we evaluate the summation over $j$ first by using Eq.~(\ref{eq:L41}), which also leads to a double sum form for $I_{D2}$. We observe substantial cancellations among the obtained double sums of $I_{D1}$ and $I_{D2}$. With the remaining sums evaluated by the formulas in Appendix~\ref{app:1}, we arrive at a closed-form expression of $I_{D}$ as shown in~(\ref{eq:IDf}), where the coefficients are listed in Table~\ref{t:ID}.
\begin{eqnarray}\label{eq:IDf}
\!\!\!\!\!\!\!\!\!I_{D}&=&\frac{m}{8(2m+2\alpha+1)^4}\Big(d_0+d_{1}\psi_{0}(m+\alpha+1)+d_{2}\psi_{0}(m+2\alpha+1)+d_{3}\psi_{0}(2m+2\alpha+1)\nonumber\\
&&+d_{4}\big(\psi_{0}(m+2\alpha+1)-\psi_{0}(2m+2\alpha+1)\big)\big(\psi_{0}(m+2\alpha+1)-\psi_{0}(2m+2\alpha+1)\nonumber\\
&&+\psi_{0}(m+\alpha+1)\big)+d_{5}\psi_{0}^{2}(m+\alpha+1)+d_{6}\big(\psi_{1}(m+2\alpha+1)-\psi_{1}(2m+2\alpha+1)\big)\Big).
\end{eqnarray}

Finally, inserting Eqs.~(\ref{eq:IAf}),~(\ref{eq:IBCf}),~(\ref{eq:IDf}) into Eq.~(\ref{eq:Tvi}), one observes cancellations of all but three terms
\begin{eqnarray}\label{eq:Tva}
\mathbb{V}\!_{h}\!\left[T\right]&=&m(m+2\alpha+1)\psi_{0}(m+\alpha+1)+\frac{m(m+2\alpha+1)}{2}\psi_{0}^{2}(m+\alpha+1)\nonumber\\
&&+\frac{m(m+2\alpha+1)}{4(2m+2\alpha+1)}\big(5m^2+5m+10\alpha m+4\alpha^2+4\alpha+2\big)\psi_{1}(m+\alpha+1).
\end{eqnarray}
Upon specializing the above expression with the $\alpha$ value in Eq.~(\ref{eq:aBH}) establishes the induced variance formula~(\ref{eq:Tv}). This completes the proof of the main result~(\ref{eq:vNv}).

\section{Summary and outlook}\label{sec:con}
As an important step towards quantifying the statistical performance of bipartite systems, we derived the exact variance of von Neumann entanglement entropy over the Bures-Hall measure in this work. The result is based on recent progress in understanding the correlation functions of the Bures-Hall random matrix ensemble.

Although the Bures-Hall ensemble attains a more involved functional form, the expressions of its first two moments turn out simpler than the ones over the Hilbert-Schmidt ensemble. Further understanding of this counterintuitive fact requires the higher order moments of von Neumann entropy of both ensembles. The results may as well help prove the conjectured Gaussian limit for large dimensional quantum systems. Future work also includes the study of other performance indicators relevant for quantum information processing, such as the fidelity, over the Bures-Hall measure.

\begin{acknowledgments}
The author thanks Shi-Hao Li and Peter Forrester for correspondence. The author also wishes to thank Santosh Kumar for providing the simulation codes.
\end{acknowledgments}

\appendix

\section{List of summation identities}\label{app:1}
In this Appendix, we list the closed-form~(\ref{eq:A1})--(\ref{eq:A5}) and semi-closed-from~(\ref{eq:A6})--(\ref{eq:A8}) finite sum identities useful in simplifying the summations in Sec.~\ref{sec:IABCD}. The identities~(\ref{eq:A1})--(\ref{eq:A3}) and~(\ref{eq:A4})--(\ref{eq:A8}) can be found in Ref.~\cite{Wei20} and Ref.~\cite{Milgram17}, respectively. Note that it is sufficient to assume $a,b\ge0, a\neq b$ in Eqs.~(\ref{eq:A1})--(\ref{eq:A4}),~(\ref{eq:A6}) and $a>m$ in Eqs.~(\ref{eq:A7}),~(\ref{eq:A8}).

\begin{flalign}\label{eq:A1}
\sum_{k=1}^{m}\psi_{0}(k+a)=(m+a)\psi_{0}(m+a+1)-a\psi_{0}(a+1)-m &&
\end{flalign}
\begin{eqnarray}\label{eq:A2}
\!\!\!\!\!\!\!\!\!\!\!\!\!\!\!\!\!\!\!\!\!\!\!\!\!\sum_{k=1}^{m}k\psi_{0}(k+a)&=&\frac{1}{2}\left(m^2+m-a^2+a\right)\psi_{0}(m+a+1)+\frac{1}{2}(a-1)a\psi_{0}(a+1)+\frac{1}{4}m\nonumber\\
&&\times(2a-m-3)
\end{eqnarray}
\begin{eqnarray}\label{eq:A3}
\!\!\sum_{k=1}^{m}k^{2}\psi_{0}(k+a)&=&\frac{1}{6}\left(2m^3+3m^2+m+2a^3-3a^2+a\right)\psi_{0}(m+a+1)-\frac{1}{6}a\left(2a^2-3a+1\right)\nonumber\\
&&\times\psi_{0}(a+1)-\frac{1}{36}m\left(4m^2+15m-6ma+12a^2-24a+17\right)
\end{eqnarray}
\begin{flalign}\label{eq:A4}
\sum_{k=1}^{m}\frac{\psi_{0}(k+a)}{k+a}=\frac{1}{2}\left(-\psi_{0}^{2}(a+1)+\psi_{0}^{2}(m+a+1)-\psi_{1}(a+1)+\psi_{1}(m+a+1)\right) &&
\end{flalign}
\begin{flalign}\label{eq:A5}
\sum_{k=1}^{m}\frac{\psi_{0}(m+1-k)}{k}=-\psi_{0}(1)\psi_{0}(m+1)+\psi_{0}^{2}(m+1)-\psi_{1}(1)+\psi_{1}(m+1) &&
\end{flalign}
\begin{flalign}\label{eq:A6}
\sum_{k=1}^{m}\frac{\psi_{0}(k+b)}{k+a}=&-\sum_{k=1}^{m}\frac{\psi_{0}(k+a)}{k+b}+\psi_{0}(m+a+1)\psi_{0}(m+b+1)-\psi_{0}(a+1)\psi_{0}(b+1)\nonumber\\
&+\frac{1}{a-b}(\psi_{0}(m+a+1)-\psi_{0}(m+b+1)-\psi_{0}(a+1)+\psi_{0}(b+1)) &&
\end{flalign}
\begin{eqnarray}\label{eq:A7}
\!\!\!\!\!\!\!\!\!\!\!\!\!\!\!\!\!\sum_{k=1}^{m}\frac{\psi_{0}(k)}{a+1-k}&=&\sum_{k=1}^{m}\frac{\psi_{0}(k)}{k+a-m}+\frac{1}{2}\left(\psi_{0}(a-m+1)-\psi_{0}(a+1)\right)^{2}-\frac{1}{2}(\psi_{1}(a-m+1)\nonumber\\
&&-\psi_{1}(a+1))
\end{eqnarray}
\begin{flalign}\label{eq:A8}
\sum_{k=1}^{m}\frac{\psi_{0}(a+1-k)}{k}=&-\sum_{k=1}^{m}\frac{\psi_{0}(k+a-m)}{k}+\left(\psi_{0}(m+1)-\psi_{0}(1)\right)\left(\psi_{0}(a-m)+\psi(a+1)\right)&&\nonumber\\
&+\frac{1}{2}\left((\psi_{0}(a-m)-\psi_{0}(a+1))^2+\psi_{1}(a+1)-\psi_{1}(a-m)\right) &&
\end{flalign}

\section{Coefficient lists of $I_{A}$, $I_{B}$, $I_{C}$, and $I_{D}$}\label{app:2}
\begingroup
\squeezetable
\begin{longtable}[!h]{rl}
\caption{Coefficients of $I_{A}$ in Eq.~(\ref{eq:IAf})}\\
\hline\hline\rule{0pt}{4ex}
$a_{0}=$ & $-18\alpha m(m+\alpha)(m+2\alpha)(m+2\alpha+1)(2m+2\alpha+1)^2\left(5m^2+10\alpha m+5m+4\alpha^2+4\alpha+2\right)$ \\
$a_{1}=$ & $-\alpha m(m+\alpha)(m+2\alpha)\big(1756m^5+8760\alpha m^4+4464m^4+15900\alpha^{2}m^3+16152\alpha m^3+3941m^3+12736\alpha^{3}m^2+19320\alpha^{2}m^2$\\
&$+9288\alpha m^{2}+1370m^2+4032\alpha^4 m+8112\alpha^3 m+5604\alpha^{2}m+1500\alpha m+147m+192\alpha^5+480\alpha^4+320\alpha^3-60\alpha-14\big)$\\
$a_{2}=$ & $-18m(m+\alpha)(m+2\alpha+1)(2m+2\alpha+1)^{2}(3m+4\alpha)\left(5m^2+10\alpha m+5m+4\alpha^2+4\alpha+2\right)$ \\
$a_{3}=$ & $12(m+2\alpha)(2m+2\alpha+1)^2\big(15m^5-30\alpha^{2}m^4+60\alpha m^4+30m^4-120\alpha^{3}m^3+27\alpha^2 m^3+72\alpha m^3+21m^3-154\alpha^{4}m^2$ \\
&$-75\alpha^{3}m^2+19\alpha^{2}m^2+24\alpha m^2+6m^2-68\alpha^{5}m-50\alpha^{4}m-16\alpha^{3}m-\alpha^{2}m-4\alpha^6+4\alpha^5+\alpha^4-\alpha^3\big)$ \\
$a_{4}=$ & $6(m+\alpha)(m+2\alpha)(2m+2\alpha+1)^{2}\big(15m^4-120\alpha^{2}m^3+30\alpha  m^3+30m^3-360\alpha^{3}m^2-228\alpha^{2}m^2+12\alpha m^2+21m^2$ \\
&$-256\alpha^{4}m-336\alpha^{3}m-176\alpha^{2}m-18\alpha m+6m-16\alpha^5-32\alpha^4-68\alpha^3-52\alpha^2-12\alpha\big)$ \\
$a_{5}=$ & $6(m+2\alpha)(2m+2\alpha+1)\big(106\alpha m^6-60m^6+636\alpha^{2}m^5-65\alpha m^5-150m^5+1506\alpha^{3}m^4+503\alpha^{2}m^4-324\alpha m^4-144m^4$ \\
&$+1784\alpha^{4}m^3+1300\alpha^{3}m^3+60\alpha^{2}m^3-235\alpha m^3-66m^3+1080\alpha^{5}m^2+1120\alpha^{4}m^2+468\alpha^{3}m^2+15\alpha^{2}m^2-58\alpha m^2$ \\
&$-12m^2+288\alpha^{6}m+320\alpha^{5}m+216\alpha^{4}m+96\alpha^{3}m+16\alpha^{2}m+16\alpha^7-8\alpha^6-12\alpha^5+2\alpha^4+2\alpha^3\big)$ \\
$a_{6}=$ & $6(m+2\alpha+1)(2m+2\alpha+1)\big(212\alpha m^6-30m^6+1512\alpha^{2}m^5+198\alpha m^5-45m^5+4212\alpha^{3}m^4+1820\alpha^{2}m^4+111\alpha m^4$ \\
&$-27m^4+5760\alpha^{4}m^3+4344\alpha^{3}m^3+1238\alpha^{2}m^3+31\alpha m^3-6m^3+3936\alpha^{5}m^2+4304\alpha^{4}m^2+2296\alpha^{3}m^2+440\alpha^{2}m^2$ \\
&$+6\alpha m^2+1152\alpha^{6}m+1680\alpha^{5}m+1480\alpha^{4}m+584\alpha^{3}m+72 \alpha^{2}m+64\alpha^7+128\alpha^6+272\alpha^5+208\alpha^4+48\alpha^3\big)$ \\
$a_{7}=$&$-12\alpha\big(212m^8+1908\alpha m^7+636m^7+7252\alpha^{2} m^6+4920\alpha m^6+823m^6+15148\alpha^{3}m^5+15660\alpha^{2}m^5+5545\alpha m^5+601m^5$\\
&$+18888\alpha^{4}m^4+26400\alpha^{3}m^4+14738\alpha^{2}m^4+3552\alpha m^4+255m^4+14192\alpha^{5}m^3+25104\alpha^{4}m^3+19608\alpha^{3}m^3+7724\alpha^{2}m^3$\\
&$+1329\alpha m^3+59m^3+6080\alpha^{6}m^2+13056\alpha^{5} m^2+13480\alpha^{4}m^2+7696\alpha^{3}m^2+2250\alpha^{2}m^2+272\alpha m^2+6m^2+1248\alpha^{7}m$\\
&$+3168\alpha^{6}m+4304\alpha^{5}m+3432\alpha^{4}m+1494\alpha^{3}m+316\alpha^{2}m+24\alpha m+64\alpha^8+192\alpha^7+416\alpha^6+512\alpha^5+324\alpha^4+100\alpha^3$\\
&$+12\alpha^2\big)$\\
\\\hline\hline
\label{t:IA}
\end{longtable}
\endgroup

\begingroup
\squeezetable
\begin{longtable}[!h]{rl}
\caption{Coefficients of $I_{B}+I_{C}$ in Eq.~(\ref{eq:IBCf})}\\
\hline\hline\rule{0pt}{4ex}
$b_{0}+c_{0}=$ & $18\alpha m(m+\alpha)(m+\alpha+1)(m+2\alpha)(m+2\alpha+1)(2m+2\alpha+1)^{3}\left(5m^2+10\alpha m+5m+4\alpha^2+4\alpha+2\right)$\\
$b_{1}+c_{1}=$ & $-2\alpha m(m+\alpha)(m+\alpha+1)(m+2\alpha)\big(1756m^6+10516 \alpha m^5+5504m^5+24660\alpha^{2}m^4+25608\alpha m^4+6479m^4$\\
&$+28636\alpha^{3}m^3+44304\alpha^{2}m^3+22151\alpha m^3+3480m^3+16768\alpha^{4}m^2+34376\alpha^{3}m^2+25380\alpha^{2}m^2+7802\alpha m^2$\\
&$+805m^2+4224\alpha^{5}m+10752\alpha^{4}m+10268\alpha^{3}m+4464\alpha^{2}m+819\alpha m+37m+192\alpha^6+576\alpha^5+560\alpha^4+160\alpha^3$\\
&$-60\alpha^2-44\alpha-7\big)$\\
$b_{2}+c_{2}=$ & $-18m(m+\alpha)(m+\alpha+1)(m+2\alpha+1)(2m+2\alpha+1)^{3}(3m+4\alpha)\left(5m^2+10\alpha m+5m+4\alpha^2+4\alpha+2\right)$\\
$b_{3}+c_{3}=$ & $12(m+\alpha+1)(m+2\alpha)(2m+2\alpha+1)^{3}\big(15m^5-30\alpha^{2}m^4+60\alpha m^4+30 m^4-120\alpha^{3}m^3+27\alpha^{2}m^3+72\alpha m^3+21m^3$\\
&$-154\alpha^{4}m^2-75\alpha^{3}m^2+19\alpha^{2}m^2+24\alpha m^2+6m^2-68\alpha^{5}m-50\alpha^{4} m-16\alpha^{3}m-\alpha^{2}m-4\alpha^6+4\alpha^5+\alpha^4-\alpha^3\big)$\\
$b_{4}+c_{4}=$ & $6(m+\alpha)(m+\alpha+1)(m+2\alpha)(2m+2\alpha+1)^{3}\big(15m^4-120\alpha^2 m^3+30\alpha m^3+30m^3-360\alpha^{3}m^2-228\alpha^{2}m^2$\\
&$+12\alpha m^2+21m^2-256\alpha^{4}m-336\alpha^{3}m-176\alpha^{2}m-18\alpha m+6m-16\alpha^5-32\alpha^4-68\alpha^3-52\alpha^2-12\alpha\big)$\\
$b_{5}+c_{5}=$ & $12(m+\alpha+1)(m+2\alpha)(2m+2\alpha+1)\big(106\alpha m^7-60m^7+742\alpha^2 m^6-141\alpha m^6-180m^6+2142\alpha^{3}m^5+378\alpha^{2}m^5$\\
&$-670\alpha m^5-219m^5+3290\alpha^{4}m^4+1743\alpha^{3}m^4-698\alpha^{2}m^4-679\alpha  m^4-138m^4+2864\alpha^{5}m^3+2448\alpha^{4}m^3$\\
&$+104\alpha^{3}m^3-613\alpha^{2}m^3-291\alpha m^3-45m^3+1368\alpha^{6}m^2+1524\alpha^{5}m^2+500\alpha^{4}m^2-84\alpha^{3}m^2-132\alpha^{2}m^2-47\alpha m^2$\\
&$-6m^2+304\alpha^{7}m+360\alpha^{6}m+172\alpha^{5}m+62\alpha^{4}m+18\alpha^{3}m+2\alpha^{2}m+16\alpha^8-16\alpha^6-4\alpha^5+3\alpha^4+\alpha^3\big)$\\
$b_{6}+c_{6}=$ & $6(2m+2\alpha+1)\big(424\alpha m^9-60m^9+4720\alpha^{2}m^8+1036\alpha m^8-240m^8+22640\alpha^{3}m^7+13764\alpha^{2}m^7+716\alpha m^7$\\
&$-399m^7+61184\alpha^{4}m^6+62444\alpha^3 m^6+17280\alpha^{2}m^6-80\alpha m^6-357m^6+102120\alpha^{5}m^5+148488\alpha^{4}m^5+74852\alpha^{3}m^5$\\
&$+13467\alpha^{2}m^5-265\alpha m^5-183m^5+108240 \alpha^{6}m^4+206232\alpha^{5}m^4+152936\alpha^{4}m^4+53828\alpha^{3}m^4+7650\alpha^{2}m^4$\\
&$-74\alpha m^4-51m^4+71744\alpha^{7}m^3+170208\alpha^6 m^3+169160\alpha^{5}m^3+90024\alpha^{4}m^3+25492\alpha^{3}m^3+3057\alpha^{2}m^3+13\alpha m^3$\\
&$-6m^3+27776\alpha^{8}m^2+79328\alpha^{7}m^2+100928\alpha^{6}m^2+74144\alpha^{5}m^2+32140\alpha^{4}m^2+7480\alpha^{3}m^2+722\alpha^{2}m^2+6\alpha  m^2$\\
&$+5248\alpha^{9}m+17664\alpha^{8}m+28608\alpha^{7}m+28512\alpha^{6}m+17544\alpha^{5}m+6216\alpha^{4}m+1112\alpha^{3}m+72\alpha^{2}m+256\alpha^{10}$\\
&$+1024\alpha^9+2432\alpha^8+3712\alpha^7+3344\alpha^6+1696\alpha^5+448\alpha^4+48\alpha^3\big)$\\
$b_{7}+c_{7}=$ & $-12\alpha(2m+2\alpha+1)\big(212m^9+2120\alpha m^8+758m^8+9160\alpha^{2}m^7+6726\alpha m^7+1162m^7+22400\alpha^{3}m^6+25294\alpha^{2}m^6$\\
&$+9200\alpha m^6+1049m^6+34036\alpha^{4}m^5+52450\alpha^{3}m^5+29970\alpha^{2}m^5+7475\alpha m^5+631m^5+33080\alpha^{5}m^4+65124\alpha^{4}m^4$\\
&$+51812\alpha^{3}m^4+20908\alpha^{2}m^4+4038 \alpha m^4+251m^4+20272\alpha^{6}m^3+48896\alpha^{5}m^3+50800\alpha^{4}m^3+29310\alpha^{3}m^3$\\
&$+9392\alpha^{2}m^3+1417\alpha m^3+59m^3+7328\alpha^{7}m^2+21056\alpha^{6}m^2+27624\alpha^{5}m^2+21476\alpha^{4}m^2+10018\alpha^{3}m^2+2550\alpha^{2}m^2$\\
&$+284\alpha m^2+6m^2+1312\alpha^{8}m+4416\alpha^{7} m+7312\alpha^{6}m+7576\alpha^{5}m+4866\alpha^{4}m+1802\alpha^{3}m+340\alpha^{2}m+24\alpha m$\\
&$+64\alpha^9+256\alpha^8+608\alpha^7+928\alpha^6+836\alpha^5+424\alpha^4+112\alpha^3+12\alpha^2\big)$\\
$b_{8}+c_{8}=$ & $-18\alpha m(m+\alpha)(m+\alpha+1)(m+2\alpha)(m+2\alpha+1)(2m+2\alpha+1)^{2}\left(7m^2+14\alpha m+7m+8\alpha^2+8\alpha+2\right)$\\
$b_{9}+c_{9}=$ & $36\alpha m(m+\alpha)(m+\alpha+1)(m+2\alpha)(m+2\alpha+1)(2m+2\alpha+1)^{2}\big(10m^3+30\alpha m^2+9m^2+28\alpha^{2}m+8\alpha^3+4\alpha^2$\\
&$+16\alpha m+3m\big)$\\
$b_{10}+c_{10}=$ & $-72\alpha m(m+\alpha)(m+\alpha+1)(m+2\alpha)(m+2\alpha+1)(2m+2\alpha+1)^{2}\big(10m^3+30\alpha m^2+12m^2+28\alpha^{2}m+22\alpha m$\\
&$+6m+8\alpha^3+8\alpha^2+4\alpha+1\big)$\\
$b_{11}+c_{11}=$ & $18\alpha m(m+\alpha)(m+\alpha+1)(m+2\alpha)(m+2\alpha+1)(2m+2\alpha+1)^{2}\big(10m^3+30\alpha m^2+3m^2+28\alpha^{2}m+4\alpha m-3m$\\
&$+8\alpha^3-4\alpha^2-8\alpha-2\big)$\\
\\\hline\hline
\label{t:IBC}
\end{longtable}
\endgroup

\begingroup
\squeezetable
\begin{longtable}[!h]{rl}
\caption{Coefficients of $I_{D}$ in Eq.~(\ref{eq:IDf})}\\
\hline\hline\rule{0pt}{4ex}
$d_{0}=$ & $m\big(36m^4+136\alpha m^3+68m^3+196\alpha^{2}m^2+188\alpha m^2+31m^2+128\alpha^{3}m+184\alpha^{2}m+64\alpha m-6m+32\alpha^4+64\alpha^3+36\alpha^2$\\
&$-4\alpha-5\big)$\\
$d_{1}=$ & $2m(2m+2\alpha+1)\left(14m^3+46\alpha m^2+29m^2+48\alpha^{2}m+60\alpha m+20m+16\alpha^3+32\alpha^2+22\alpha+5\right)$\\
$d_{2}=$ & $4(2m+2\alpha+1)\big(32\alpha^4+64\alpha^3+48\alpha^2+16\alpha+30m^4+126\alpha m^3+69m^3+192\alpha^{2}m^2+204\alpha m^2+56m^2+128\alpha^{3}m$\\
&$+200\alpha^{2}m+106\alpha m+19m+2\big)$\\
$d_{3}=$ & $-4\big(60m^5+312\alpha m^4+168m^4+636\alpha^{2}m^3+672\alpha m^3+181m^3+640 \alpha^{3}m^2+1000\alpha^{2}m^2+528\alpha m^2+94m^2+320\alpha^{4}m$\\
&$+656\alpha^{3}m+508\alpha^{2}m+176\alpha m+23m+64\alpha^5+160\alpha^4+160\alpha^3+80\alpha^2+20\alpha+2\big)$\\
$d_{4}=$ & $8(m+2\alpha+1)(2m+2\alpha+1)^{2}\left(3m^2+6\alpha m+3m+4\alpha^2+4\alpha+1\right)$\\
$d_{5}=$ & $-2m(m+2\alpha+1)^{2}(2m+2\alpha+1)^{2}$\\
$d_{6}=$ & $4(m+2\alpha+1)(2m+2\alpha+1)^{3}\left(5m^2+10\alpha m+5m+4\alpha^2+4\alpha+2\right)$\\
\\\hline\hline
\label{t:ID}
\end{longtable}
\endgroup

\bibliography{BHv}
\end{document}